\documentclass[11pt]{article}
\usepackage{moriond,epsfig}

\def\be{\begin{equation}}
\def\ee{\end{equation}}
\def\bea{\begin{eqnarray}}
\def\eea{\end{eqnarray}}

\def \gr {\kern 2pt\hbox{}^\circ{\kern -2pt K}} 

\def \Om {\Omega}

\def\be{\begin{equation}}
\def\ee{\end{equation}}
\def\bea{\begin{eqnarray}}
\def\eea{\end{eqnarray}}

\def\si{\sigma}

\input epsf.tex
\begin{document}
\vspace*{4cm}

\title{COSMOLOGICAL IMPLICATIONS OF GALAXY CLUSTERS:
\\ 
BEST-FIT MODELS}

\author{V.N.~LUKASH}

\address{Astro Space Centre of P.N.Lebedev Physical Institute, Moscow}

\maketitle\abstracts{
The galaxy cluster power spectrum and mass/temperature functions are currently 
the most precise observational tools for constraining the theory of the 
formation of large scale structure (LSS) in the Universe. Complementing these 
tests by the observational data at larger (cosmic microwave backgroud anisortopy
(CMBA)) and smaller (distribution of $Ly_\alpha$ clouds) scales opens the 
way to a straightforward determination of the cosmological parameters in 
simplest dark matter (DM) models. We argue that such a 'minimal data set' is 
free from systematic effects and can indicate quite precisely the parameters 
of spatially flat mixed DM model with a positive cosmological constant and no 
cosmic gravitational waves.
}

\section{Introduction}
Three lessons from the galaxy cluster physics are often implied in 
building-up the true cosmology directly from observations: 
\begin{itemize}
\item{the cluster abundance test (the famous $\sigma_8$ argument or 
cluster mass/temperature function from data at $z\sim 0$)},
\item{the cluster power spectrum (the distribution of clusters in space)},
\item{the cluster evolution test (the cluster abundance {\it{vs}} redshift)}.
\end{itemize}

While the latter still requires better statistics and cluster mass determination, 
the first two are considered today as the most precise cosmological tests
related straightforwardly to the linear density perturbation field which gave 
the birth to LSS in the Universe (see the review papers at this Conference).
 
A good concordance of the power spectrum of galaxy clusters in the range 
$k\in (0.03, 0.2)h/$Mpc reconstructed from different optical and 
X-ray data (e.g. the talk of P.Schuecker at this Conference and refs therein) 
evidences a low level of possible systematic uncertainties in cluster data
and encourages to apply the $\chi^2$ minimization method for the restoration 
of the cosmological parameters in simplest DM models.
 
Certainly, few additional tests have to be taken into account to cover a
broader range of scales and, thus, to improve the stability of the method and 
perform a more comprehensive analysis. However, the danger may arise here in 
possible introduction of the unclear systematics brought by additional tests, 
that often makes the result of the analysis dependent on a set of the selected data.

\section{Minimal Data Set}

The determination of cosmological parameters from some geometrical tests and 
numerous LSS observations has been carried out in many papers. Though the results 
are always stable for the prediction of a small matter abundance ($\Omega_m < 1)$, 
the concrete values of $\Omega_m$ and other cosmic parameters leave something to be 
desired depending strongly on both a family of the theoretical models chosen 
(the number of free parameters) and a set of observations taken for the analysis. 
Under such a situation it is tempting to form the current 'minimal data set' suffering 
less from the systematic effects, and to see if the selected data are mutually in 
agreement implying the same cosmological parameters.
 
The modern status of observational cosmology prompts to clearly indicate at least 
three scale regions which have to be taken into account for the problem to be solvable:
\begin{itemize}
\item{very large scales ($\sim 1000h^{-1}$Mpc)},
\item{LSS scales ($\sim 10-100h^{-1}$Mpc)}, and
\item{small (sub-megaparsec) scales}.
\end{itemize}
Fortunately, all these three levels can be suggested today by the 'most trustable' 
available data related to the primordial density perturbation field throgh simple 
linear integrals: 
\begin{itemize}
\item{the first group -- by COBE 4-year data}, 
\item{the second -- by a group of LSS data selected below}, and
\item{the third -- by data on the power and distribution of $Ly_\alpha$ clouds}. 
\end{itemize}

If the first and third groups are monopolic (few observational points in each group) 
destining just to stabilize the spectrum amplitudes in the asymptotic scale regions, 
then the central (second) group is the most important for the analysis allowing for a
subtle 'fine tuning' of the result. This group contains a bulk of observational points 
determining both the spectral slopes and amplitudes within the LSS scales. 
Novosyadlyj {\it et al}.~\cite{n} suggested the following seven data points (in the second group)
 which meet common 
agreements in literature and can be thought of as 'most trustable' and accurate:
\begin{itemize}
\item{the $\sigma_8 - \Omega_m$ relation found by cluster abundance at 
$z\sim 0$ (one point)},
\item{the mean peculiar velocity of galaxies in a sphear of radius 
$50h^{-1}$ Mpc (one point)},
\item{the Abell-ACO cluster power spectrum in scales $k\in (0.03, 0.2)h$/Mpc 
(three effective degrees of freedom)},
\item{the amplitude and position of the first acoustic peak in $\Delta T/T$ spectrum
(two points)}.
\end{itemize}
These data points have been shown to be self-consistent (excluding some points from 
the searching procedure results only in a change of the best-fit parameter values 
within the range of their corresponding standard errors, see Sections 3,4). 
More of that, some other data (e.g. the evolution of cluster number density, see
Section 4) can be added to these seven points with a no change of the result. 

Below we present the cosmological models and constrain the parameters using only this 
'minimal data set'. The reported analysis does not include observations of distant 
supernovae, galaxy power spectra, a bulk of the CMB data (except the two mentioned 
points for the location and position of the first acoustic peak and the COBE 
normalization), the local measurements of Hubble constant, the age of globular 
clusters, the evolution tests and some other data. It is interesting to see if and how 
our result confronts with these 'other' observational tests mentioned.

\section{The Model and Determination of Its Parameters}

The usual cosmological paradigm -- a scale free power spectrum of scalar
primordial perturbations which evolve in a multicomponent medium to
form the large scale structure of the Universe -- is compatible with
the observed LSS and CMBA. 
Most inflationary scenarios predict a scale free primordial power spectra
of scalar density fluctuations ($P(k)\sim k^{n}$ with arbitrary $n$) as
well as gravity waves which contribute to the power spectrum of CMB
temperature fluctuations $({\Delta T\over T})_{\ell}$ at low spherical
harmonics. But models with a minimal number of free parameters, such
as the scale invariant ($n=1$) standard cold DM model or the standard 
mixed (cold + hot) DM model are excluded by observational data. 

Better agreement between theory and observations 
can be achieved in models with a larger number of parameters: 
cold dark matter (CDM) or mixed dark matter (MDM) with baryons, 
a tilted primordial power spectra, spatial curvature, cosmological constant, 
and a tensor contribution to the CMB anisotropy power spectrum. However, 
to ensure a better stability and convergence of the minimization method 
on the basis of the 'minimal data set', the total number of free parameters 
in the model should not be large (currently, not exceeding ten).

One way to improve the standand MDM is an introduction of cosmic gravity waves (CGW). 
However, the direct approach (just introducing the tensor mode while keeping the 
$\Lambda$-term vanishing) is not effective (see Mikheeva {\it et al}.~\cite{m}): 
the best-fit model normalized by $\sigma_8$ and COBE data requires too much hot matter 
($\Omega_\nu \ge 0.2$) and too low first acoustic peak (inconsistent with $Ly_\alpha$ 
and CMBA observations). Another way -- the introduction of cosmological constant 
without CGW -- appears more powerful.

Novosyadlyj {\it et al.}\cite{n} have performed such an analysis for $\Lambda$MDM 
spatially flat cosmological model ($\Omega_{\Lambda}+\Omega_m=1$) without tensor mode, 
also neglecting a possible effect of the early reionization which could reduce the amplitude 
of the first acoustic peak in the CMB anisotropy spectrum. 
The reason for the restriction of flat models is strongly motivated by  
the results from the BOOMERANG experiment ~\cite{boo}, 
whereas neglecting the CGW and reionization is mainly technical: 
the extraction of these parameters in models with non-zero $\Lambda$-term would require 
more accurate experimental data than those available today.

Among the free cosmological parameters one is discrete (the number of species of 
massive neutrinos) and six are continuous within the following ranges:
\begin{itemize}
\item{$N_\nu=1,2,3\;$, the number of massive neutrino species},
\item{$n\in (0.7, 1.4)$, the tilt of primordial spectrum},
\item{$\Omega_m\in (0, 1)$, the abundance of total matter},
\item{$\Omega_\nu\in (0, 0.4)$, the abundance of hot matter},
\item{$\Omega_b\in (0, 0.2)$, the abundance of baryons},
\item{$h\in (0.4, 0.8)$, the Hubble constant in units 
$100\rm{km}\rm{s}^{-1}\rm{Mpc}^{-1}$},
\item{$b_{Cl}\in (1, 5)$, the bias of galaxy cluster distribution}.
\end{itemize}
The abundance of cold dark matter and cosmological $\Lambda$-term are found from the
equations $\Omega_{c}=\Omega_m-\Omega_\nu-\Omega_b$ and $\Omega_\Lambda=1-\Omega_m$, 
respectively; the bias of the cluster power spectrum with respect to the dark matter 
distribution is supposed to be linear and scale independent in the range of scales considered.

The method for detecting cosmological parameters from the 'minimal data set' has been tested 
for the stability with help of the constructed mock sample of observational data. The only pourly 
determined parameter (from the LSS data alone) turned out to be the baryonic abundance. To fix 
$\Omega_b$ with better accuracy we have added a single geometrical test in our analysis: 
the Big Bang nucleosysthesis constraint, $\Omega_bh^2=0.019\pm 0.0024$. With such an addition 
the method reveals as very stable and finds all the parameters of 'true' model whenever possible.

The best-fit model with the minimum of $\chi^2$ (when all parameters are free) is presented 
in the first line of Table 1. The rest eleven lines display some other best-fit models (with 
some of the seven parameters fixed) which have got within the 
$1\sigma$ contour (from the first-line model). In all the models $\chi^2_{min}$ is in the range 
\[
N_F-\sqrt{2N_F}\le \chi^2_{min}\le N_F+\sqrt{2N_F} 
\]
which is expected for a Gaussian distribution with $N_F$ degrees of freedom (the number of 
observational points minus the number of free model parameters).

\begin{table*}[th]
\caption{Cosmological parameters of $\Lambda$MDM models without 
cosmic gravity waves
(the free/fixed parameters are given with/without standard errors)}
\begin{center}
\def\onerule{\noalign{\medskip\hrule\medskip}}
\medskip
\begin{tabular}{|ccccccccc|}
\hline
&&&&&&&&\\
No  &$N_{\nu}$     & $\chi^2_{min}$  &$n$  & $\Omega_m$&$\Omega_{\nu}$& $\Omega_b$ & $h$    & $b_{cl}$ \\ [4pt]
\hline
&&&&&&&&\\
1&1 & 4.63&1.12$\pm$0.10&0.41$\pm$0.11&0.059$\pm$0.028&0.039$\pm$0.014&0.70$\pm$0.12&2.23$\pm$0.33\\
2 &2 & 4.80&1.13$\pm$0.10&0.49$\pm$0.13&0.103$\pm$0.042&0.039$\pm$0.014&0.70$\pm$0.13&2.33$\pm$0.36\\
3 &3 & 5.07&1.13$\pm$0.10&0.56$\pm$0.14&0.132$\pm$0.053&0.040$\pm$0.015&0.69$\pm$0.13&2.45$\pm$0.37\\ [4pt]
4 &1 & 5.27&1.12$\pm$0.09&0.51$\pm$0.07&0.074$\pm$0.041&0.053$\pm$0.003&0.60&2.43$\pm$0.26 \\
5 &1 & 4.65&1.12$\pm$0.10&0.39$\pm$0.05&0.058$\pm$0.026&0.037$\pm$0.002&0.72&2.19$\pm$0.23 \\  [4pt]
6 &1 &12.23&1.07$\pm$0.09&1.00&0.116$\pm$0.086&0.118$\pm$0.027&0.40$\pm$0.05&3.15$\pm$0.39 \\
7 &2 &10.17&1.10$\pm$0.09&1.00&0.177$\pm$0.086&0.099$\pm$0.022&0.44$\pm$0.05&3.10$\pm$0.38 \\
8 &3 & 8.80&1.12$\pm$0.09&1.00&0.219$\pm$0.084&0.085$\pm$0.019&0.47$\pm$0.05&3.07$\pm$0.38 \\ [4pt]
9&1-3 & 6.54&1.04$\pm$0.10&0.30&0.000$\pm$0.005&0.038$\pm$0.013&0.71$\pm$0.12&2.25$\pm$0.19 \\[4pt]
10&1 & 6.18&1.00&0.45$\pm$0.12&0.042$\pm$0.032&0.038$\pm$0.014&0.71$\pm$0.13&2.44$\pm$0.31 \\
11&3 &10.43&1.00&1.00&0.159$\pm$0.069&0.075$\pm$0.021&0.51$\pm$0.07&3.23$\pm$0.35  \\ [4pt]
12&1-3& 6.92&1.00&0.30&0.000$\pm$0.010&0.034$\pm$0.009&0.75$\pm$0.10&2.25$\pm$0.20  \\ [4pt]

\hline
\end{tabular}
\end{center}
\end{table*}

\section{Results and Discussion}

As it is seen from Table 1, the considered observational data on LSS 
of the Universe can be explained by a flat $\Lambda$MDM inflationary model 
with a tilted spectrum of scalar perturbations and vanishing tensor contribution.
The best fit parameters are: $N_{\nu}=1$, $n=1.12\pm0.10$, $\Omega_m=0.41\pm0.11$,
$\Omega_{\nu}=0.059\pm0.028$, $\Omega_b=0.039\pm0.014$ and $h=0.70\pm0.12$.
The CDM density parameter is $\Omega_{c} = 0.31\pm0.15$ and $\Omega_{\Lambda}$ 
is considerable, $\Omega_{\Lambda}=0.59\pm0.11$.

If all parameters are free (line 1) the model with one sort of massive 
neutrino provides the best fit to the data. However, there are
only marginal differences in $\chi^2_{min}$ for $N_\nu =1,2,3$ 
(lines 1,2,3, respectively), therefore, with the given accuracy of the data 
we cannot currently conclude whether -- if massive neutrinos are present at all -- 
their number is one, two, or three. 

The spectral index is close the Harrison-Zel'dovich and coincides with the COBE
prediction. The neutrino matter density is about the baryon abundance: 
$\Omega_\nu\sim\Omega_b\sim 10\%$ of $\Omega_m$. 

Surprisingly, the prediction of model parameters from the 'minimal data set' is
consistent with observations of the nearby and distant SNIa, the age of
globular clusters, and the existence of rich galaxy clusters at $z\ge 0.5$.
However, the comparison with galactic power spectra creates a problem
since different systematics in galaxy catalogues and a 
scale-dependent bias because of non-linear clustering of galaxies at small scales.   

Notice, that the value of Hubble parameter found from the LSS 
tests is consistent with local measurements of Hubble constant.
It is interesting to see that the value of Hubble constant 
unticorrelates with the total matter abundance (lines 4,5): roughly,
the production of both factors remains constant, $\Gamma\equiv\Omega_m h\simeq 0.3$.
Note, that without hot matter ($\Lambda$CDM) $\Gamma\simeq 0.22$ (see lines 9,12) 
in concordance with other results.

Furthermore, increasing the number of massive neutrino species from 1 
to 3 leads to an increase of $\Omega_{\nu}$ from 0.06 to 0.13 and to a decrease 
of $\Omega_{\Lambda}$ from 0.59 to 0.44 (lines 1,2,3).
The correlation between $\Omega_\nu$ and $\Omega_m$ can be approximated at the
'maximum likelihood ringe' by the following equation:
\[
\Omega_{\nu}\simeq 1.3\Omega_m^2 -0.44\Omega_m +0.023
\]

There are some other interesting models staying within the $1\sigma$ range form 
the best-fit first line. Among them are the matter dominated models with zero 
$\Lambda$-term (lines 6,7,8): all these models require rather high abundance of the 
hot matter (up to $22\%$ for three sorts of massive neutrino) and extremely low 
value of the Hubble constant ($h<0.5$) which is in obvious disagreement with the 
local SNIa measurements. 

The  models with low $\Omega_m \sim 0.3$ (lines 9,12) fit the observational data somewhat 
less good than the best model ($\Delta\chi^2_{min}\simeq 2$) but all predictions 
are still within the $1\sigma$ contour. Such model prefers a high Hubble parameter
($h{}^{>}_\sim 0.7$) and no massive neutrinos. Obviosly, it is the standard $\Lambda$CDM.

Concerning the perfectly scale invariant primordial power spectrum (lines 10,11,12), 
these models prefer $\Lambda$MDM (with a somewhat lower neutrino content than in 
the best-fit model in line 1) if the remaining parameters are initially free (line 10),
and the $\Lambda$CDM if the matter content is initially fixed as low (line 12).
As for the matter dominated model with $n=1$ (line 11), it is just the standard MDM 
with a low Hubble parameter, $h\simeq 0.5$, practically coinciding with the line 8.

If the hot component is eliminated from the very beginning or $\Omega_{\nu}$ is fixed at 
the small value defined by the lower limit of the neutrino mass
$\sqrt{\delta m_{\nu}^2}=0.07$ from the Super-Kamiokande experiment, 
$\Omega_{\nu}=7.4\times 10^{-4}N_{\nu}/h^2$, we obtain the best-fit value for the matter 
density parameter $\Omega_m\simeq 0.39\pm0.11$ and the Hubble constant 
$h=0.62\pm0.12$.

The errors in the best fit parameters presented in Table 1 are
the square roots of the  diagonal elements of the covariance
matrix.  More information about the accuracy of the determination of 
parameters and their sensitivity to the data used can be obtained from the 
contours of confidence levels presented in Fig. 1.  
These contours show the confidence regions which contain 68.3\% (solid line),
95.4\% (dashed line) and 99.73\% (dotted line) of the total probability
distribution in the two dimensional sections of the six-dimensional
parameter space of $\Lambda$MDM models, if the probability distribution is Gaussian.

As one can see in Fig.1a  the iso-$\chi^2$ surface is rather prolate from
the low-$\Omega_m$ - high-$n$  corner to high-$\Omega_m$ - low-$n$.
This indicates some degeneracy in $n-\Omega_m$ parameter plane, which
can be expressed by the following equation which roughly describes the
'maximum likelihood ridge' in this plane within the $1\si$:
\[
n\sqrt{\Omega_m}\simeq 0.73
\]
A similar degeneracy in the $\Omega_{\nu}- \Omega_m$ plane
in the range $0\le\Omega_{\nu}\le 0.17$, $\;0.25\le\Omega_m \le 0.6$
(Fig.1c) was already discussed above. The both degeneracies have clear physical
explanation. The rest contours are quasi-spherical and consistent with the 
Gaussian character of the errors.

\begin{figure*}[tp]
\epsfxsize=16truecm
\epsfbox{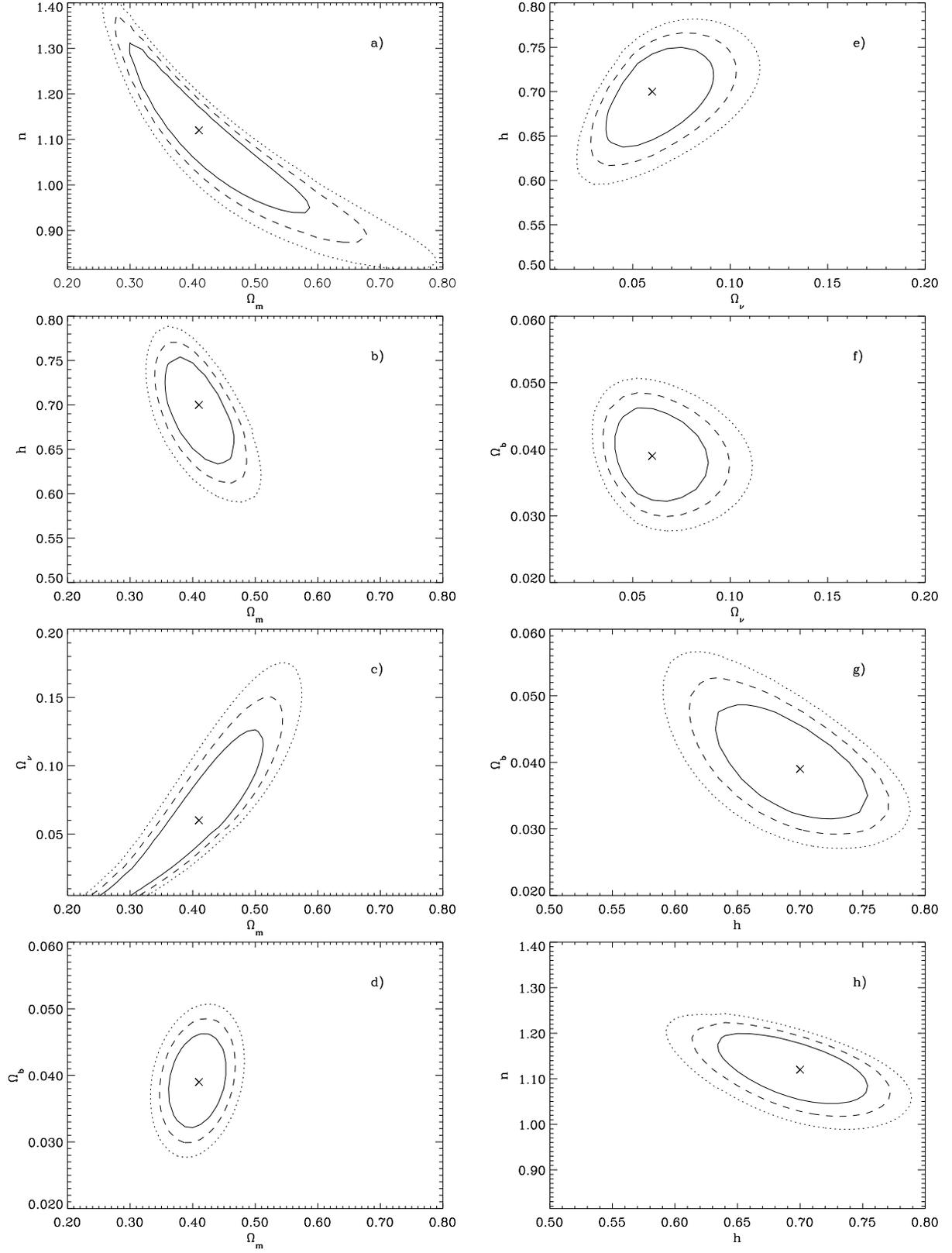}
\caption{The likelihood contours (solid line - 68.3\%, dashed - 95.4\%, dotted
- 99.73\%) of the tilted $\Lambda$MDM model with $N_{\nu}=1$ and
parameters from Table 1 (line 1) in the different planes of
$n-\Omega_m-\Omega_{\nu}-\Omega_b-h$ space.
 The parameters not shown in a given diagram are set to their best fit values.}
\label{Lcm1}
\end{figure*}

One important question is how each point of the data influences the final result. 
To estimate this we have excluded some data points from the searching procedure.  
Excluding any part of observable data results only in a slight change of the 
best-fit values of $n$, $\Omega_m$, and $h$ within the ranges of their 
corresponding standard errors. This indicates that the data are mutually in
agreement implying the same cosmological parameters (within the error-bars).
Concerning the parameter $\Omega_\nu\;$, a very important is the small scale constraint: 
the $Ly_\alpha$ tests reduce the hot dark matter content from $\Omega_{\nu}\sim 0.22$ 
to $\sim 0.07$. The rest tests stabilize the value of $\Omega_\nu$ within its
still considerable errorbars.  

The most crucial test for the baryon content is of course the nucleosynthesis
constraint (LSS alone does not determine $\Omega_b$). Its $\sim 6\%-1\sigma$-accuracy 
safely keeps $\Omega_b h^2$ near its median value 0.019. The parameter $\Om_b$ in turn 
is only known to $\sim 36\%$ accuracy due to the large errors of other experimental 
data (reduced to the uncertainty in Hubble constant).
The obtained accuracy of $h$ ($\sim17\%$ from the 'minimal data set') is 
{\it{better}} than the one assumed from direct local measurements ($\sim 23\%$ only).

Thus, all the data points used in the analysis are non-contradictory and mutually 
self-consistent: all they are important for searching the best-fit cosmological parameters.

\begin{figure*}[tp]
\epsfxsize=14truecm
\epsfbox{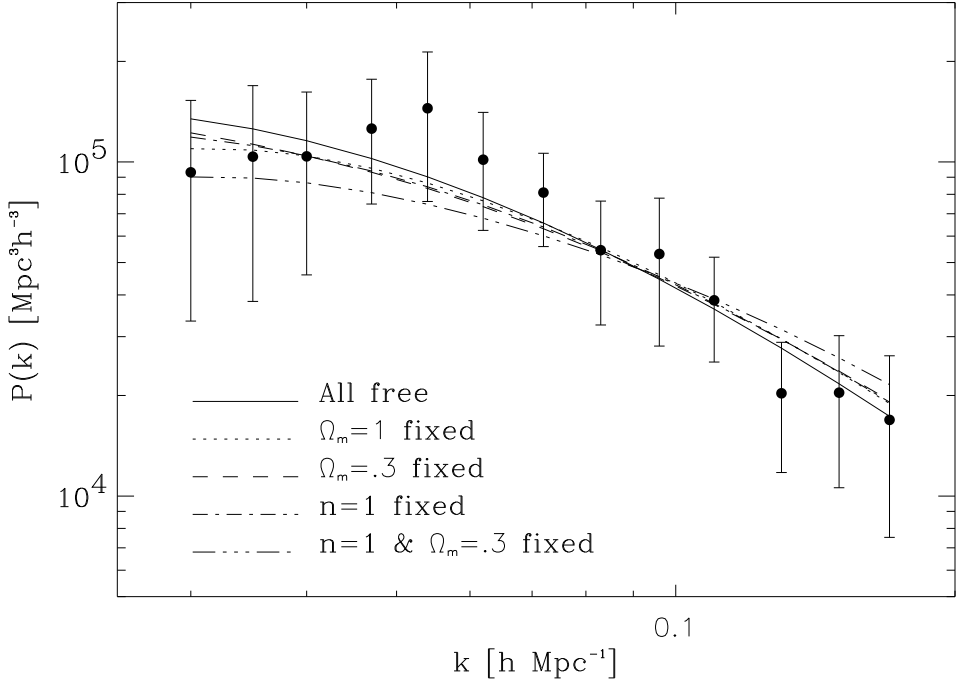}
\caption{The observed Abell-ACO power spectrum (filled circles) and the
theoretical spectra predicted by tilted $\Lambda$MDM models with parameters
taken from Table 1 ($N_\nu =1$).}
\label{Lcm1}
\end{figure*} 

Finally, let us emphasize an important spectrum property. The experimental Abell-ACO power 
spectrum and the theoretical predictions for some best-fit models are shown in Fig. 2. 
As we see the best-fit spectra do not have any pecularity at the scale $k\sim 0.05h$/Mpc.
More of that, the theoretical spectra bend evidently at larger scale! 
This can create a problem for $\Lambda$MDM: if better data on the spatial distribution
of galaxy clusters will confirm the presence of the bent/feature in the power spectrum 
the whole $\Lambda$MDM paradigm will be in trouble. 
A way out could be that the real spectrum bent occurs at larger scale, $k_{bent}\le 0.05h$/Mpc, 
as indicated by the $\Lambda$MDM models. (Some observational data hint on such a possibility, e.g.
Miller and Batuski~\cite{mb}). Then, what is the physical reason of the spectrum feature at $k\sim 0.05h$/Mpc? 
Does it have a baryonic nature (the modulation of the power spectrum by the acoustic waves existing
in the early Universe in the baryon-radiation plasma)? If so, then the fraction of baryons should be 
higher than we think today, $\frac{\Omega_b}{\Omega_m}> 10-15\%$.
All these questions only stress the importance of further cluster investigations.

\section{Conclusions}

The experimental data set with a minimal current level of systematic effects can be 
used to constrain successfully the cosmological parameters of simple dark matter models.
 
The 'minimal data set' includes {\it{(i)}} the LSS constraints based on galaxy cluster observations 
(the Abell-ACO power spectrum~\cite{ret97}, the density fluctuation amplitude $\sigma_8$ 
derived from the mass function of nearby~\cite{gir98} and distant~\cite{bah98} clusters, 
the mean peculiar velocity of galaxies in a sphere of radius $50h^{-1}$Mpc~ \cite{kol97},  
the position and amplitude of the first acoustic peak in the angular power spectrum of 
CMBA~\cite{boo}), 
{\it{(ii)}} the {\it{small-scale}} constraints on the amplitude and tilt of the power spectrum  
obtained from $Ly_\alpha$ clouds at z=2-3~ \cite{gn98,cr98}, 
{\it{(iii)}} the {\it{large-scale}} constraint on the amplitude of the power spectrum 
obtained from the COBE data~\cite{bun97}, and 
{\it{(iv)}} the Big Bang nucleosynthesis constraint~ \cite{bur99}.

The results of the determination of the cosmological parameters for spatially flat
$\Lambda$MDM model without cosmic gravity waves by the $\chi^2$ minimization
method are listed as follows:

\begin{itemize}
\item{The $\Lambda$MDM model with the best-fit parameters from the first line in Table 1 matches the observational data set best: 
\\
$\;\;\;\;*\;$ one sort of massive neutrino is slightly preferable, however the data do not distiguish
models with one, two or three massive neutrino species within $1\sigma$ c.l.;
\\
$\;\;\;\;*\;$ the tilt of the power spectrum is consistent with the Harrison-Zel'dovich one;
\\
$\;\;\;\;*\;$ the abundance of matter and cosmological constant is roughly half-to-half, 
\\
$\;\;\;\;*\;$ the abundance of hot matter is close to the abundance of baryons and consists $10-15\%$ 
of the total matter abundance, 
\\
$\;\;\;\;*\;$ the Hubble constant fits the local SNIa measurements.}

\item{Fixing a low $\Omega_m =0.3$ the $\Lambda$CDM model without hot matter and with high $h{}^>_\sim 0.7$
matches the observational data set best.} 

\item{Fixing a high $\Omega_m =1$ the standard MDM model with high abundance of hot matter 
(up to 
$\Omega_\nu =0.22$) and extremely low $h\le 0.5$ matches the observational data set best. The standard CDM/MDM 
models with $h > 0.5$ are ruled out at very high confidence level, $99.99/95\%$ c.l. respectively.}

\item{Raising fixed Hubble parameter decreases the total matter abundance keeping the production of 
both factors approximately constant: $\Gamma\equiv\Omega_m h\simeq 0.3$. 
The same procedure for a fixed 
low $\Omega_m$ universe ($\Lambda$CDM) keeps the shape parameter at the level $\Gamma\simeq0.22$.}

\item{Increasing the number of massive neutrino species raises the abundances of both the hot and total matter, 
the corresponding correlation between $\Omega_{\nu}$ and $\Omega_m$ (when the rest parameters are fixed to 
their best-fit values) is approximated by the equation:
\\
 $\Omega_{\nu}\simeq 1.3\Omega_m^2-0.44\Omega_m+0.023$.}

\item{Increasing the tilt of power spectrum raises the value of $\Lambda$-term: $n\sqrt{\Omega_m}\simeq 0.73$.}

\item{For all best-fit models 
\\
$\;\;\;\;*\;$ the biasing parameter of rich galaxy clusters remains in the range 
$b_{Cl}\simeq 2.2-3.3$, the $1\sigma$ confidence interval is $1.5\le b_{Cl}\le 3.5$;
\\ 
$\;\;\;\;*\;$ the power spectra of matter density perturbations do not have percularity at $k\sim 0.05h$/Mpc (a better understanding of the cluster power spectrum break or baryonic feature at $k\sim 0.05h$/Mpc is required).}
 
\item{The best-fit models constrained by the 'minimal data set' are 
\\
$\;\;\;\;*\;$ consistent with the age of globular clusters, observations of nearby and distant SNIa, 
evolution of the comoving number density of clusters with redshift;
\\ 
$\;\;\;\;*\;$ inconsistent with galactic power spectra (a better understanding of the 
non-linear biasing and different systematics of galaxy catalogues is required).}

\item{The $\Lambda$-term is strongly indicated by the 'minimal data set' based on cluster observations.
However, still a low accuracy of the present observational data does not allow 
\\
$\;\;\;\;*\;$ to constrain the set of cosmological parameters sufficiently,
\\ 
$\;\;\;\;*\;$ to discriminate between the $\Lambda$MDM and $\Lambda$CDM models (even at $1\sigma$ c.l.).} 

\end{itemize}

It may well be that the DM nature is more complex than just the $\Lambda$MDM without CGW. 
The progress in both, the LSS theory and observations based on
galaxy clusters helps to solve the DM problem, the key problem of the cosmological model.

\section*{Acknowledgments} This work was partly supported by INTAS (97-1192) and RFBR (01-02-16274a).
The author is grateful to the Organising Committee for the hospitality and support.


\section*{References}
\begin{thebibliography}{2}

\bibitem{n}
B. Novosyadlyj, R. Durrer, S. Gottloeber, V.N. Lukash, S. Apunevich, A\&A {\bf 356}, 418 (2000).

\bibitem{m} 
E.V. Miheeva, V.N. Lukash, N.A. Arkhipova, A.M. Malinovsky, Russian Astron. J. {\bf 45}, 195 (2001).

\bibitem{boo} 
A. Melchiorri {\it et al.}, ApJ {\bf 536}, L63 (2000).

\bibitem{mb}
C.J. Miller, D.J. Batuski, ApJ in press (2001).

\bibitem{ret97}
J. Retzlaff {\it et al.}, New Astronomy {\bf 3}, 631 (1998).

\bibitem{gir98}
M. Girardi {\it et al.}, ApJ {\bf 506}, 45 (1998).

\bibitem{bah98}
N.A. Bahcall, X. Fan, ApJ {\bf 504}, 1 (1998).

\bibitem{kol97}
T. Kolatt, A. Dekel, ApJ {\bf 479}, 592 (1997).

\bibitem{gn98}
N.Y. Gnedin, MNRAS {\bf 299}, 392 (1998).

\bibitem{cr98}
R.A.C. Croft {\it et al.}, ApJ {\bf 495}, 44 (1998).

\bibitem{bun97}
E.F. Bunn, M. White, ApJ {\bf 480}, 6 (1997).

\bibitem{bur99}
S. Burles, K.M. Nollett, J.N. Truran, M.S. Turner, Phys.Rev.Lett. {\bf 82}, 4176 (1999).



\end{thebibliography}
\end{document}